\documentclass[12pt]{iopart}
\pdfoutput=1
\usepackage{graphicx}
\usepackage{cite}
\usepackage{amsfonts}
\usepackage{wrapfig}

\begin{document}

\title[Explosive Ising]
      {Explosive Ising}
\author{Sebastian Angst$^1$, Silvio R. Dahmen$^2$, Haye Hinrichsen$^3$, \\Alfred Hucht$^1$, and Martin~P.~Magiera$^1$}
\address{$^1$ Fakult\"at f\"ur Physik, Universit\"at Duisburg-Essen, 47048 Duisburg, Germany}
\address{$^2$ Instituto de Fisica, Universidade Federal do Rio Grande do Sul, \\ \ \ \ 91501-970 Porto Alegre RS, Brazil}
\address{$^3$ Universit\"at W\"urzburg, Fakult\"at f\"ur Physik und Astronomie, \\ \ \ \ 97074 W\"urzburg, Germany}

\begin{abstract}
We study a two-dimensional kinetic Ising model with Swendsen-Wang dynamics, replacing the usual percolation on top of Ising clusters by explosive percolation. The model exhibits a reversible first-order phase transition with hysteresis. Surprisingly, at one of the transition flanks the global bond density seems to be equal to the percolation threshold.
\end{abstract}

\vspace{-6mm}
\ead{sebastian.angst@uni-due.de}
\ead{silvio.dahmen@ufrgs.br}
\ead{hinrichsen@physik.uni-wuerzburg.de}
\ead{fred@thp.uni-due.de}
\ead{martin.magiera@uni-due.de}

\parskip 2mm 


\noindent
Three years ago Achlioptas \textit{et al.} discovered that a slight modification of standard percolation on a complete graph turns the continuous transition into an abrupt one~\cite{Achlioptas}. This surprising phenomenon, marketed as ``explosive percolation'', attracted a lot of attention and was studied on scale-free networks~\cite{Fortunato,Cho1}, regular two-dimensional lattices~\cite{Ziff}, and in various other situations~\cite{Moreira,Cho2,Cho3,Friedman,Manna,Souza,Rozenfeld,Araujo}. Initially the explosive transition was believed to be discontinuous, but later studies revealed that the transition is extremely sudden but actually continuous\cite{daCosta,Grassberger,Riordan,Park,TianShi}. 

While in ordinary bond percolation each bond is set independently with probability~$p$, the Achlioptas process uses a dynamical procedure, where the bonds are placed in sequence. To this end one randomly selects two pairs of neighboring lattice sites which are not yet connected by a bond. Depending on the actual bond configuration, a weight is assigned to each of the pairs by multiplying the sizes of the clusters to which the two lattice sites belong (or squaring the size if both sites belong to the same cluster). If the weights of the two pairs are different, the pair with the \textit{lower} weight is connected by a new bond. Otherwise, if the weights are equal, one of the pairs is randomly chosen and connected. This process is repeated until the density of bonds exceeds the value of the control parameter $p$.

\begin{figure}
\centering\includegraphics[width=90mm]{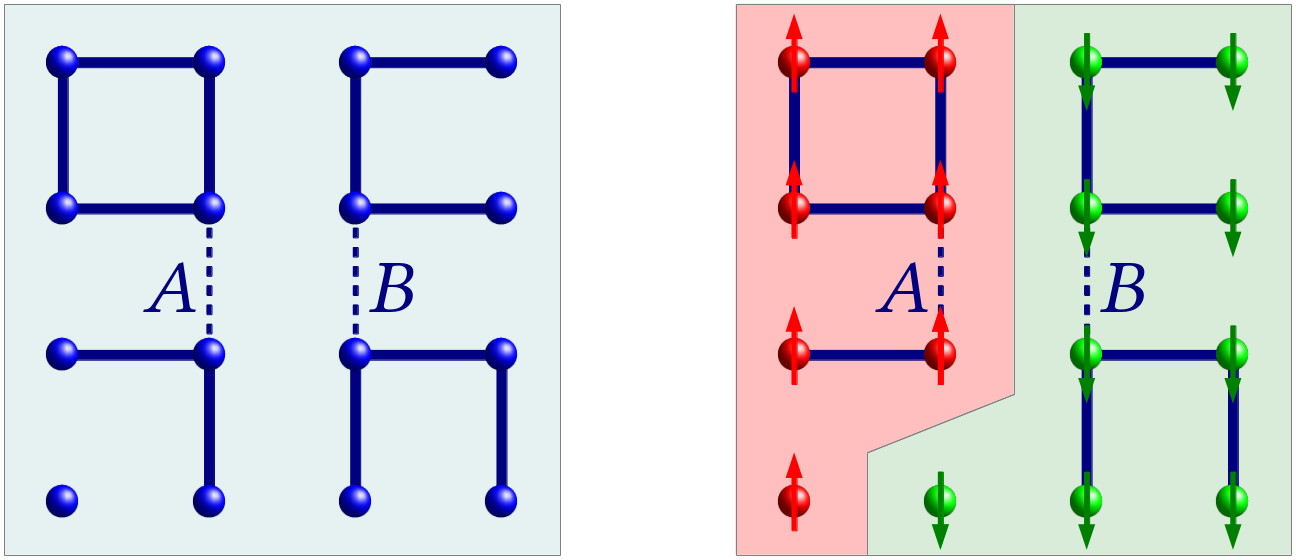}
\caption{\label{fig:experc}\small Left: Dynamics of explosive bond percolation on a square lattice. Two vacant bonds $A,B$ (shown as dashed lines) are randomly selected. As bond $A$ would connect two clusters of sizes 2 and 4, its weight is $2\cdot 4=8$, whereas bond $B$ has the weight $3 \cdot 3=9$. The bond with the \textit{lower} weight is set and the procedure is repeated until the density of bonds exceeds the parameter~$p$. Right: ``Explosive'' Ising model with two spin domains (red and green). Here a Swendsen-Wang update uses the same dynamical rule for placing bonds but only within domains of equally oriented spins. }
\end{figure}

As the dynamic rule is defined in such a way that new connections are preferentially added between \textit{small} clusters, it is plausible that the transition point is shifted to higher values of the percolation probability. For example, on a two-dimensional square lattice, where the transition of ordinary bond percolation takes place at $p^{\rm\tiny perc}_c=1/2$, the  Achlioptas-dynamics described above shifts the transition point to $p^{\rm\tiny expl}_c = 0.526562(3)$~\cite{Grassberger}. Surprisingly, the transition is not only shifted, but it changes also qualitatively, exhibiting a sudden transition, where a large percolation cluster is formed. This is why the transition is called ``explosive''.

In this Letter we demonstrate that the concept of explosive cluster growth can also be applied successfully to kinetic spin models with cluster dynamics. Here we will focus on a particularly simple case, namely, the two-dimensional Ising model with Swendsen-Wang dynamics~\cite{SwendsenWang}. As will be shown below, the use of an explosive cluster dynamics turns the continuous Ising transition in a discontinuous one, preserving the $Z_2$-symmetry of the model. As this modification of the dynamics is expected to break detailed balance, the stationary state of such an explosive Ising model will no longer be an equilibrium state.\\

\noindent\textbf{Definition of the model:}\\
Before defining the explosive Ising model let us briefly recall the Swendsen-Wang (SW) algorithm applied to the Ising model~\cite{SwendsenWang}. The SW cluster algorithm is a dynamical update rule which works as follows. At first percolation clusters are grown \textit{within} the actual spin clusters. For each of these clusters one chooses a random  number $\pm 1$ which is then assigned to all its spins. More specifically, in case of the Ising model the SW update for a given configuration of spins consists of the following steps:
\begin{itemize}
\item[(i)] Remove all bonds.
\item[(ii)] Set all bonds between equally oriented nearest-neighbor spins with probability $p$.
\item[(iii)] Identify all clusters of sites which are connected by bonds.
\item[(iv)] For each of these clusters generate a random number $\pm 1$ with equal probability and assign it to all its spins.
\end{itemize}
It was proven that this dynamics evolves into the equilibrium state of the Ising model without critical-slowing-down and that the percolation probability is related to the temperature by $p=1-e^{-2J/k_BT}$. 

Let us now modify the SW dynamics of the Ising model by replacing ordinary with explosive percolation.  Since bonds can no longer be distributed independently, one first has to determine the number of potential bonds $n$. This results into the following ``explosive'' Swendsen-Wang dynamics for the Ising model (see right panel of Fig.~\ref{fig:experc}): 
\begin{enumerate}
\item Clear all bonds.
\item Count the number $n$ of links between neighboring spins of the same orientation.
\item Select two vacant bonds between equally oriented spins and determine their weight according to the product rule in the same way as in explosive percolation.
\item Place a bond at the link with the lower weight (if the weights are equal select one of them randomly).
\item Repeat (iii) and (iv) until the number of bonds exceeds $np$, where $p\in[0,1]$ is the control parameter of the model.
\item Finally, as in ordinary SW dynamics, assign a random spin orientation to each of the clusters.
\end{enumerate}
%
%
\begin{figure}[b]
\centering\includegraphics[width=150mm]{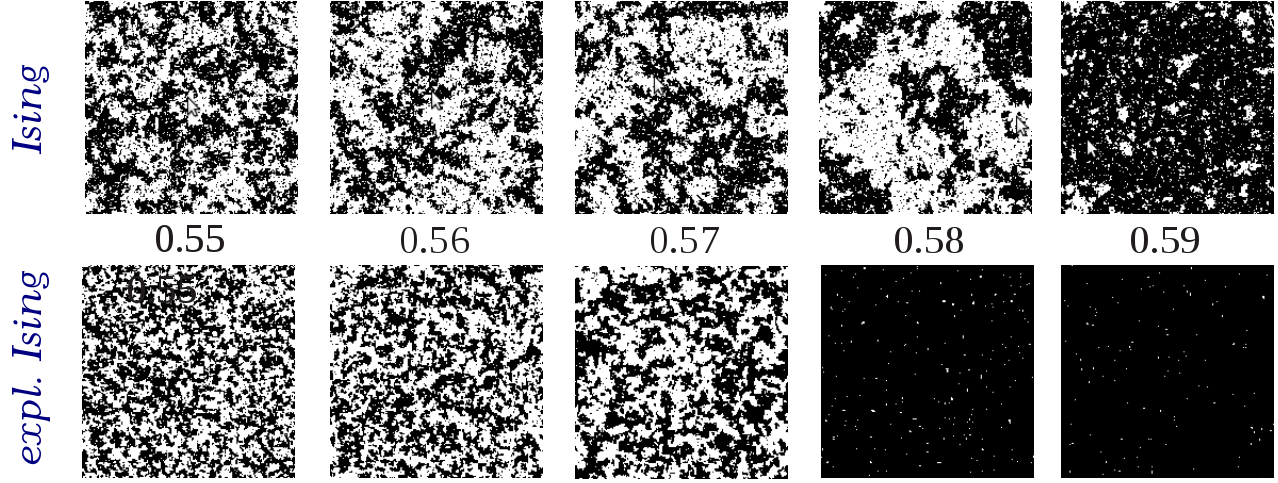}
\caption{\small\label{fig:demo} Typical snapshots of spin configurations for the ordinary and the explosive Ising model with $p$ increasing from 0.55 to 0.59.}
\end{figure}
%
Fig.~\ref{fig:demo} shows typical spin configurations of the ordinary Ising model compared with those of the explosive variant defined above, increasing the parameter $p$ from 0.55 to 0.59. For small values of $p$ both models are in the disordered phase, although the domains in the explosive variant seem to be somewhat smaller. As $p$ is increased, the Ising model displays a continuous transition at $p_c^{\rm\tiny Ising}= 2-\sqrt{2} \approx 0.585786$ into a partially ordered state, while the explosive variant switches suddenly to an almost completely ordered state at some value between 0.57 and 0.58 (see below). \\

\begin{figure}
\centering\includegraphics[width=100mm]{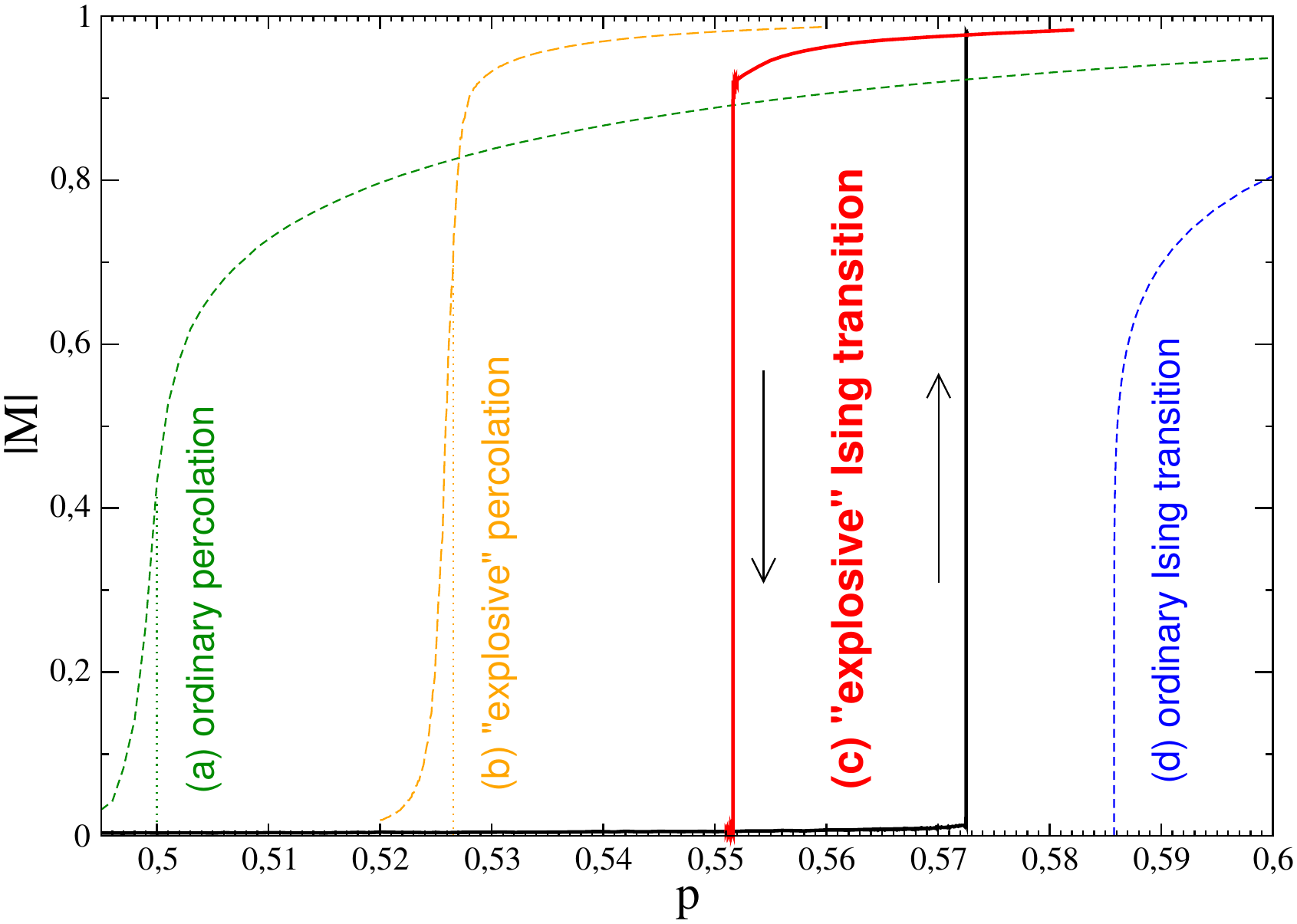}
\caption{\small \label{fig:hysteresis}Order parameter as a function of the percolation probability $p$ for various models discussed in the text. From left to right: (a) Average density of the largest cluster in ordinary bond percolation on a $3000 \times 3000$ lattice and periodic boundary conditions. The dotted line indicates the transition point in the limit $L \to \infty$. (b) Largest cluster size in explosive percolation as defined above on a periodic $500 \times 500$ lattice. (c) Hysteresis of the order parameter $|M|$ in the explosive Ising model on a square lattice with $512 \times 512$ sites and periodic boundary conditions. (d) Same data for the ordinary two-dimensional Ising model  (analytical result).}
\end{figure}

\noindent\textbf{Order parameter hysteresis:}\\
In explosive percolation the process of adding new bonds is not reversible, i.e. in a single realization the density of bonds can only be increased. In the present model, however, the explosive cluster dynamics is part of a continually repeating update procedure, allowing one to increase and decrease the percolation probability while the model evolves in time. 

Since a spin model with cluster dynamics switches frequently from positive to negative magnetization in the ordered phase, the appropriate order parameter is the absolute value of the magnetization $M$. The solid lines in Fig.~\ref{fig:hysteresis} show how $|M|$ varies with the control parameter $p$ on a lattice with $N=512 \times 512$ sites and periodic boundary conditions. As can be seen, the transition seems to be discontinuous, exhibiting a pronounced hysteresis with the flanks located at the values $p_{c,1}$ and $p_{c,2}$ (see Table 1).
\begin{table}[b]
\begin{center}
\begin{small}\begin{tabular}{|c|c|c|c|c|} \hline
ordinary & explosive & explosive Ising & explosive Ising & ordinary\\[-1mm]
percolation & percolation & left flank & right flank & Ising\\ \hline
$p^{\rm\tiny perc}_c=1/2$ & $p^{\rm\tiny expl}_c=0.526562(3)$ & $p_{c,1} = 0.5516(3)$ & $p_{c,2}= 0.5725(3)$ & $p_c^{\rm\tiny Ising}= 0.585786$ \\ \hline
\end{tabular}
\caption{Transition points for various models discussed in the text.}
\label{table}
\end{small}\end{center}
\end{table}

\newpage\noindent
The hysteresis may be explained as follows. For $p \approx 0.5$ the Achlioptas selection rule generates smaller clusters than in the usual Ising model. Since the percolation clusters are confined to domains of the same spin orientation, their growth is suppressed like in a finite system with open boundary conditions, explaining why nothing happens at the usual threshold $p^{\rm\tiny expl}_c \approx 0.527$ of explosive percolation. However, as the size of the Ising domains grows with $p$, one eventually reaches a point $p_{c,2}$  where this stabilization mechanism breaks down. Here one observes the immediate formation of a system-spanning cluster, self-sustained from inside by the as from now super-critical Achlioptas process. Lowering $p$ again the system first remains in the ordered state until the explosive percolation transition is reached from above. As the percolation process takes place on a slightly porous support, the transition point is already reached at a value $p_{c,1}$ which is somewhat higher than the usual transition point of explosive percolation  $p^{\rm\tiny expl}_c$.

\begin{figure}
\centering\includegraphics[width=160mm]{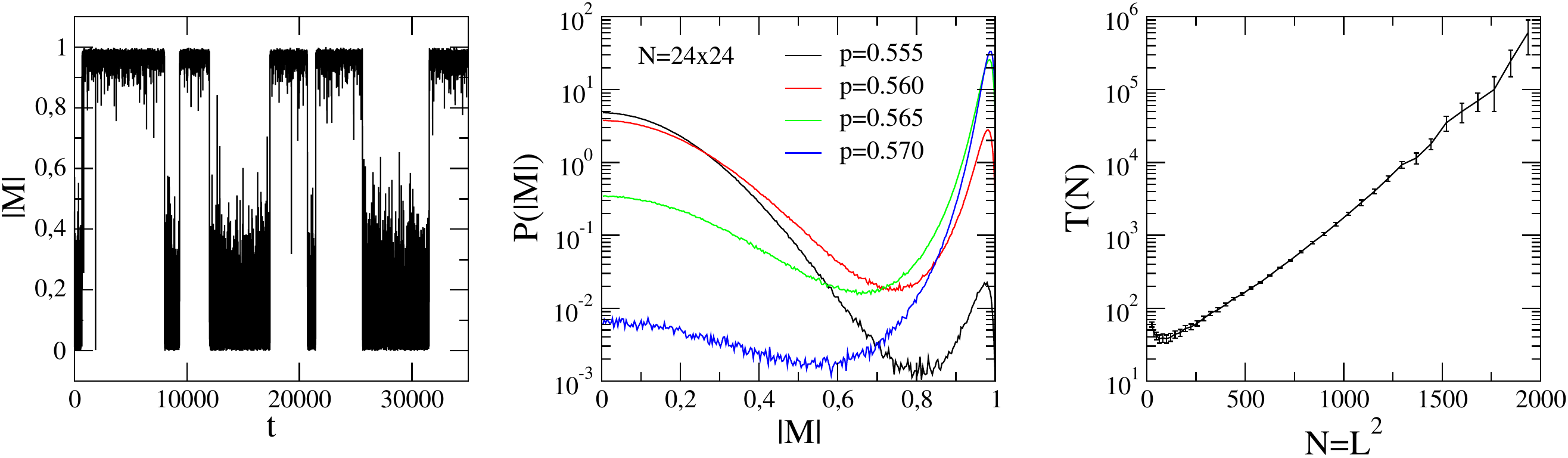}
\caption{\label{fig:flipping}\small Left: Spontaneous flipping of the order parameter $|M|$ in a system with $24\times 24$ sites inside the hysteresis loop at $p=0.563$. Middle: Distribution of the order parameter averaged over long time for various values of $p$ in a $24\times 24$ system. Right: The average time $T(N)$ between two flips at $p=0.563$ grows exponentially with the system size.}
\end{figure}

\noindent\textbf{Spontaneous flipping and order parameter distribution:}\\
As usual in systems with a first-order phase transition, we find that the order parameter in finite systems flips occasionally between the disordered and the ordered phase (left panel of Fig.~\ref{fig:flipping}). As expected, the corresponding probability distributions of the order parameter (see central panel) show two maxima separated by a valley, whose relative sizes depend on $p$. We also verified that the flipping time $T(N)$ diverges exponentially with the number of sites $N=L^2$, meaning that  the ordered and the disordered phase are both thermodynamically stable in the coexistence region $p_{c,1} <p<p_{c,2}$.

\noindent\textbf{Density of equally oriented neighbors at the transition flanks:}\\
%
Let $b$ be the total number of bonds generated in the update procedure. We finally present the surprising conjecture that the global bond density 
\begin{equation}
q = \frac {b}{2N}
\end{equation}
is exactly equal to $p^{\rm\tiny expl}_c$ at the left transition flank in the thermodynamic limit. 

\noindent
Remarkably, a similar relation holds at the transition point of the \textit{ordinary} Ising model on a square lattice~\cite{BondDensity}. To see this let us consider the number $n$ of equally oriented neighboring spins, which is related to the internal energy $E=-J\sum_{\langle i,j \rangle} s_i s_j$ by
\begin{equation}
n \;=\; - \frac{E+E_0}{2J}\,,
\end{equation}
where $E_0=-2NJ$ the lowest possible energy where all spins are parallel. As bonds are randomly set between equally oriented spins with probability $p$, we have $b \approx n p$ so that the global bond density is given by
\begin{equation}
q \;=\; \lim_{N \to \infty}\, \frac{n p}{2N}\,.
\end{equation}
Since the internal energy of the Ising model on a square lattice at the critical point $p_c^{\rm\tiny Ising}=1-e^{-2J/k_BT_c}=2-\sqrt{2}$ is known to be $E_c = -  \sqrt{2} N J$, we find that
\begin{equation}
q_c \;=\; \frac{n_c p_c^{\rm\tiny Ising}}{2N} \;=\; \frac{\sqrt{2}+2}{4} \, (2-\sqrt{2}) \;=\; \frac12\,.
\end{equation}
This means that the transition in the ordinary Ising model with Swendsen-Wang dynamics takes place when the global bond density reaches the value $1/2$ -- the same value as the critical threshold for percolation \textit{without} Ising spins. The actual transition point $p_c^{\rm\tiny Ising}\approx 0.5858$ is of course larger since bonds between different spin clusters cannot be set so that an enhanced probability is needed to reach a global bond density of $1/2$.

\begin{figure}
\centering\includegraphics[width=120mm]{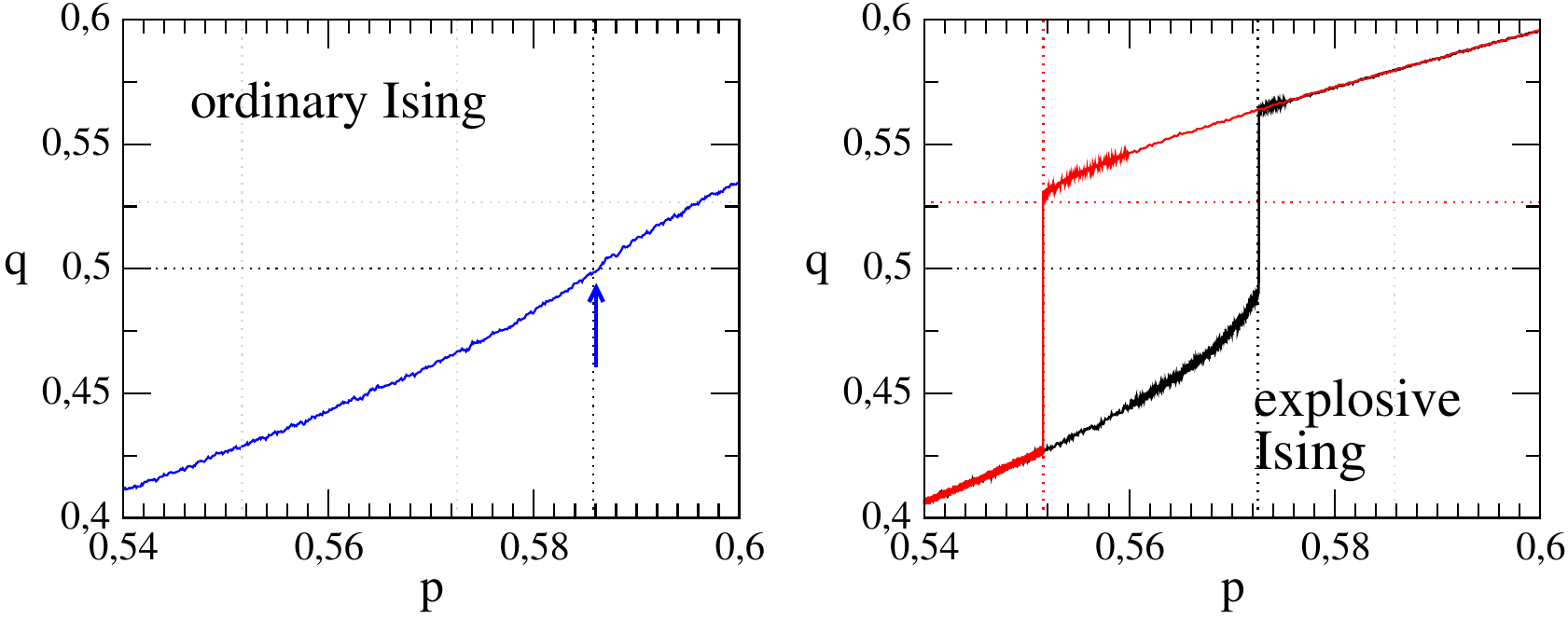}
\caption{\label{fig:bonddens}\small Left: Global bond density $q$ set during the SW update as a function of the percolation probability $p$ in the ordinary Ising model (left) and the explosive variant (right). The horizontal dotted lines indicate where $q=p_c^{\rm\tiny perc}=1/2$ and $q=p_c^{\rm\tiny expl}=0.526562(3)$.
}
\end{figure}

Surprisingly, we find numerically that the same relation holds at the left transition flank of the explosive Ising model, where the bond density $q_{c,1}$ reaches the value $p_c^{\rm\tiny expl}$ of explosive percolation without Ising spins (see Fig.~\ref{fig:bonddens}). However, at the right flank the transition occurs at a lower value $q_{c,2} \approx 0.49$ which may tend to $1/2$ in the limit of very large lattices.

\newpage
\noindent\textbf{Concluding remarks:}\\
In this Letter we introduced and studied an explosive variant of the Ising model, arriving at two conjectures:
\begin{itemize}
\item Even though the Achlioptas-process was found to be continuous, we think that the transition in the explosive Ising model is genuinely discontinuous because it exhibits a very clear hysteresis.
\item The global density of bonds in the SW update at the left transition flank of the hysteresis is conjectured to be equal to the threshold $p_c^{\rm\tiny expl}$ of explosive percolation without Ising spins. An analogous relation holds already for the usual Ising model. At the right transition flank the bond density is lower, probably close to $1/2$. 
\end{itemize}
It would be interesting to explore these conjectures more deeply and to apply similar ideas to other spin systems with cluster dynamics. 

\noindent\textbf{Acknowledgments}\\
This work was supported financially by the German Academic Exchange Service (DAAD) under the Joint Brazil-Germany Cooperation Program PROBRAL.

\section*{References}


\begin{thebibliography}{99}
\bibitem{BondDensity}
Although this results should be already known, we could not identify an explicit reference.
\end{thebibliography}
\end{document}